\documentclass[preprint,proceedings]{rmaa}

\usepackage{rmaacite}

\SetYear{2002}
\SetConfTitle{Winds, Bubbles \& Explosions, ed. Jane Arthur}

\title{The Dynamical Evolution of Narrow Line Regions}
\author{M.~A. Dopita, G.~V. Bicknell, R.~S. Sutherland \& C.~J. Saxton
\affil{Research School of Astronomy \& Astrophysics, 
The Australian National University.} }

\fulladdresses{
\item M.~A. Dopita, G.~V. Bicknell, R.~S. Sutherland \& C.~J. Saxton: Research 
School of Astronomy \& Astrophysics, Institute of Advanced Studies, 
The Australian National University, Cotter Road, Weston Creek, ACT 2611, 
Australia (Michael.Dopita@anu.edu.au).}

\shortauthor{Dopita}
\shorttitle{dynamical Evolution of the NLR}

\addkeyword{ISM:hydrodynamics}
\addkeyword{ISM: Jets and outflows}
\addkeyword{Galaxies:AGN}
\addkeyword{Galaxies:Seyfert}
\addkeyword{Galaxies:evolution}
 
\abstract{
We review recent progress by the RSAA team in elucidating the dynamical
evolution of the various classes of Narrow Line Regions (NLR) in active 
galaxies. }

\listofauthors{M.~A. Dopita, G.~V. Bicknell, R.~S. Sutherland \& C.~J. 
Saxton, C.~J.}
\indexauthor{Dopita, M.~A.}
\indexauthor{Bicknell, R.~S.}
\indexauthor{Sutherland, R.~S.}
\indexauthor{Saxton, C.~J.}

\begin{document}

\maketitle



\section{NLR Zoology}

\label{sec:intro}

Optical spectroscopy and dynamical analysis has revealed that the so-called
narrow line regions (NLR) are an almost ubiquitous signature of the presence
of an active nucleus in a galaxy (AGN). The optical diagnostic diagrams of 
\scite{Baldwin81}, \scite{Veilleux87} and \scite{Osterbrock92} serve to
classify a region as an NLR, and such regions can be further sub-divided
into two general categories: Seyfert narrow line regions (NLRs) and the 
Low Exitation Nuclear Emission Line Regions (LINERs), which will not be 
discussed here

Homogenous sets of high-quality spectrophotometric data are now available
for the NLR of most classes of AGN. Notable amongst these are complete sample 
of southern elliptical galaxies by \scite{Phillips85}, the similar-sized
sample of nearby northern spirals by \scite{Ho95}, the compilations of 
\scite{Veilleux87}, and \scite{Veron00}, the Seyfert galaxy study of
Veilleux (1991a,b,c), the compact radio-luminous sample of Gelderman \&
Whittle (1994), the luminous infrared galaxy survey of \scite{Kim95},
the ultraluminous infrared galaxy survey of \scite{Kim98} and 
\scite{Veilleux99}, the warm IRAS sample of \scite{Kewley01a}, and the
radio-excess IRAS galaxies (Drake et al, 2003, in press). NLR are
found not only in nearby galaxies, but they also appear to be a ubiquitous 
feature of QSOs \scite{Francis91}, and they provide the dominant
form of near-nuclear optical emission in the distant high-power radio sources  
(\pcite{Best00a}; \pcite{Best00b}; \pcite{Inskip02a}; \pcite{Inskip02b};  
\pcite{DeBreuck2000}).

There is increasing evidence that the narrow-line regions (NLRs) associated
with many classes of active galactic nuclei (AGN) have a complex dynamical 
and excitation evolution. Evidence for a cocoon of strong, auto-ionising
and radiative shocks is particularly compelling for luminous classes 
of radio galaxies (see the theory  by \pcite{Dopita95}; \pcite{Dopita96}
and \pcite{Bicknell97}). These include the steep-spectrum  radio sources (CSS), 
\pcite{Fanti90}, unbeamed Gigahertz-peaked sources (GPS) (see the recent 
review  by \pcite{O'Dea98}), compact symmetric objects (CSO) 
(\pcite{Wilkinson94}) or compact double sources (CD) \cite{Phillips82}. 
Together, these represent an appreciable fraction (10-30\%) of luminous radio 
sources. Not only are such sources very luminous at  radio frequencies, 
but they also are very luminous in optical emission lines, and spectra by 
\scite{Gelderman94} reveal intense ``narrow line'' emission 
with line ratios similar to those of Seyfert 2 galaxies.  For these 
objects, the line emission scales with radio power, and the continuity of 
properties across these different classes of sources argues strongly 
that the kinetic energy supplied by the radio-emitting jets may 
provide a substantial fraction of the power radiated at other
wavelengths by shocked gas associated with the NLRs of these galaxies.

Evidence for excitation by boths shock and photons has  been adduced from
the powerful high-redshift radio galaxies. The study by Best et al. (2000a,b) 
revealed that for the 3C radio galaxies ($z\sim 1$) radio lobes small,
they are predominantly shock-excited, but when the radio lobes have burst out 
into intergalactic space the ionised gas left behind is predominantly 
photoionized. Similar results were obtained by \scite{DeBreuck2000} for the 
very high redshift radio galaxies.

The Seyfert galaxies are generally found in spiral galaxy hosts, often display
lines of high-excitation, and evidence of non-gravitational motions,
but are radio-quiet. Dynamical signatures of strong shocks are apparent 
in only $5-10\%$ of cases (eg.  \pcite{Whittle96}). Many of these found
amongst the more radio-luminous galaxies including Mrk78 (\pcite{Pedlar89}), 
NGC2992  \scite{Allen99} or Mrk 1066 \scite{Bower95}. The weak collimated 
radio-jets frequently seen in such objects often show close correlations 
between the radio and the optical morphology (\pcite{Allen99}; \pcite{Axon98}; 
\pcite{Bower95}; \pcite{Capetti95}; \pcite{Falcke98}; \pcite{Haniff88}; 
\pcite{Whittle88}). Power requirements are  modest, typically $10^{41}- 10^{44}$
~ergs~s$^{-1}$  {\it c.f} the luminous radio sources 
($10^{45} - 10^{46}$~erg~s$^{-1}$). The remainder of Seyfert galaxies appear to 
be photoionised \cite{Evans99}.

The Infrared Astronomical Satellite (IRAS) revealed a large population of
galaxies with intense circum-nuclear star formation and which emit the bulk 
of their radiation in the infrared. Many of these have $\log(L_{FIR})>11.0$ 
and are referred to as luminous infrared galaxies (LIRGs). Although 
it is clear that the IR luminosity derives from dust reprocessing of 
other sources of luminosity in the galaxy, the nature of the nuclear source 
is still in debate. Most of the ultraluminous sources (ULIRGS; {\it eg} 
\pcite{Goldader95}) and the majority of lower luminosity galaxies 
\scite{Kim95} are star formation dominated. Indeed, \scite{Condon91} 
concluded that the far infrared luminosity and radio properties of LIRGs 
can be explained entirely by compact nuclear starburst events. However, 
\scite{Sanders88} have argued that LIRGs contain a dust enshrouded AGN and 
Vielleux, Sanders \& Kim (1997) have used near IR and optical spectroscopy 
to search for broad emission lines, indicative of AGN, which they found in 
some 25-30\% of ULIGs of their sample, and Lonsdale, Smith \& Lonsdale (1993)
concluded that Active Galactic Nuclei (AGN) are the dominant powering mechanism.
It is most likely that both mechanisms contribute to the overall energy output, 
and where each provides a similar flux, the nuclear spectra
are ``composite''. Such spectra are particulary common in luminous IR galaxy
samples. They have been identified by  \scite{Kim95}, \scite{Kim98} and 
\scite{Veilleux99},  the studies of \scite{Veron97}, \scite{Concalves99}
and in the extensive survey of warm IR galaxies by \scite{Kewley01a}. 
They are also found in radio-excess IRAS galaxies, which are a rare but 
important class  of radio-intermediate sources. \scite{Drake02} has shown
that 40-45\% of such objects are compact and have steep radio spectral indices,
{\it i.e.} they fit the definition of CSS or GPS radio sources.

From all of this, it should be clear that both shocks and photoionization
from the central nucleus are important in exciting the NlR. In this review, 
we will attempt to provide observational and theoretical insight into the 
dynamical evolution of these various classes of NLR objects, and infer under 
what circumstances each of these excitation mechanisms may dominate.

\section{Models of the Dynamical Evolution of the NLR}
\label{sec:models}
\subsection{The GPS/CSS Sources}
In their initial phases of activity the relativistic jets produced by AGN 
are likely to interact strongly with the interstellar medium (ISM). 
Thus, the early evolution of the lobe is likely to
provide insight into both the nature of the circum-nuclear ISM and of the
jet itself. Modulo issues relating to the absolute power of the radio jet,
the youngest AGN are probably also the smallest, and therefore an
appropriate place to start our examination of the dynamical evolution of AGN
is the gigahertz-peak sources (GPS) and the compact symmetric sources (CSS), 
which are  likely to be triggered by the merger of a low-mass system onto
a massive and evolved Elliptical galaxy.
In what follows we draw upon the results of \scite{Bicknell02}.

In his early models of these sources \pcite{Begelman96} conjectured that the
mean pressure in the radio hot-spot of these sources is in constant ratio to
the pressure of the cocoon of hot and relativistic plasma which surrounds
the jet. If the ambient ISM declines in density as a power law of the
distance from the nucleus, $n(R)=n_{0}\left( R/R_{0}\right) ^{-\delta },$
then Begelman's conjecture is valid for $\delta \sim 2$. This conjecture has
been verified in detail (Carvalho \& O'Dea 2002a; 2002b). The model of
\scite{Bicknell97}, which allows for the expansion losses
provides the velocity of advance, $v_{\mathrm{B}}$, of the bow-shock into a
uniform medium following such a power-law with radius:
\begin{equation}
\frac{v_{\mathrm{B}}}{c}\approx 0.056\left[ \frac{F_{46}}{n_{0.01}}\right]
^{1/3}\left[ \frac{R}{\mathrm{kpc}}\right] ^{\left( \delta -2\right) /3}
\label{eqn1}
\end{equation}
where here $n_{0.01}$ is the density at $R=1$~kpc in units of 0.01~cm$^{-3}$
and the jet power $F_{46}$ is given in terms of 10$^{46}$~ergs~s$^{-1}$.

The optical emission in the CSS sources shows clear signatures of shock
excitation, with systematic velocity offsets of $300-500$~km s$^{-1}$, lines
broadened and split by $\sim 500$ km s$^{-1}$, and a strong alignment
between the radio jets and optical emission (de Vreis 1997, 1999). Whilst
such effects were predicted by the BDO model, the shocks in this model were
produced by radiative wall shocks, which would require relatively high ISM
densities. This would permit only a low velocity of advance of the bow-shock
consistent with an expansion velocity of less than 1000km s$^{-1}$ in the
walls of the cocoon, which is inconsistent with recent observations which
suggest instead 0.1-0.3 times the speed of light. Also, the observations of
de Vreis (1999) also show that the line emission is not concentrated around
the head of the radio lobe, but trails behind it in the walls.

A likely explanation is that the majority of the ISM is in fact
characterized by a low density medium with a strongly radially declining
density gradient, but with much denser clouds are embedded in it. This is
entirely consistent with what we know about the structure of the ISM in
galaxies, with a cold and/or molecular component embedded in a hot medium at
coronal temperatures, which in elliptical galaxies has a temperature of
order $(1-2)\times 10^{7}$K and $\delta \sim 3/2.$ \scite{Irwin96}. In such a
cloudy medium, the overpressure in the cocoon surrounding the jets drives
slower, radiative shocks into the dense clouds. For clouds lying outside the
jet core itself, the ratio of the velocity of advance of the jet and the
cloud shock velocity, $v_{\mathrm{S}}$ is given in terms of the cloud
density, and the intercloud density in its vicinity, by
\begin{equation}
\frac{v_{\mathrm{B}}}{v_{\mathrm{S}}}=\zeta ^{1/2}\left( \frac{n_{\mathrm{C}}
}{n_{\mathrm{IC}}}\right) ^{1/2}  \label{eqn2}
\end{equation}
where $\zeta \sim 10-100$ is the ratio of the ram pressure of the jet and
pressure in the cocoon of shocked gas surrounding it.

Cloud shocks will become radiative when the cooling timescale behind the
shock, $\tau _{\mathrm{cool}}$, is comparable to the jet expansion
dynamical timescale, $\tau _{\mathrm{dyn}}$. The cooling timescale is given
by radiative shock models (\pcite{Dopita95}, \pcite{Dopita96}), and in the 
velocity range $200-900$ km~s$^{-1}$ can be approximated as:
\begin{equation}
\tau _{\mathrm{cool}}\approx 23000n_{\mathrm{C}}^{-1}\left[ \frac{v_{\mathrm{
S}}}{300\mathrm{km~s}^{-1}}\right] ^{3.9}  \mathrm{~yrs.}  \label{eqn3}
\end{equation}
The dynamical timescale is:
\begin{equation}
\tau _{\mathrm{dyn}}=32600\left[ \frac{v_{\mathrm{B}}}{0.1c}\right]
^{-1}\left[ \frac{R}{\mathrm{kpc}}\right] \mathrm{~yrs.} \label{eqn4}
\end{equation}
Approximate equality of these timescales when combined with equation (2)
means that the cloud to intercloud density contrast has to be at least $%
10^{3}$ and possibly as large as $10^{4}$, consistent with cloud
temperatures of $10^{3}-10^{4}$K, or turbulent motions of order 10 km s$^{-1}
$.

Modelling such a large density contrast (including radiative cooling) within
a hydrodynamic code is a challenging business. \scite{Sutherland02} have 
developed a code able to tackle this based on the VH-1 code of the 
Virginia group ({\it see} \textsf{http://wonka.physics.ncsu.edu/pub/VH-1}. 
This (originally adiabatic) PPM code has undergone numerous
changes in its Riemann solver, and it features an ``oscillation filter''
designed to supress a striping numerical instablitiy seeded by numerical 
noise, driven by directional splitting and exacerbated by radiative cooling. 
This code has recently been ported to the MPI parallel environment.

Some results of the 2D implementation of this code are shown in Figure 1,
which shows a pair of clouds being shocked by the passage of the jet near
them. When clouds are submitted to an overpressure large enough to drive
{\em non-radiative} shocks into them, the clouds are rapidly ablated and torn
apart, which entrains matter into the cocoon as a whole. However, when the
shocks are radiative, the clouds cool and are crushed to a density which
tends to put them into pressure equilibrium with the surrounding jet or
cocoon environment. This makes them much more resistant to destruction on
short timescales, considerably extending the cloud destruction time scale.
Simulations of  radio-lobe/cloud interactions are available as movies 
on \textsf{http://macnab.anu.edu.au/radiojets/gps}.

\begin{figure}
   \leavevmode
    \includegraphics[width=\columnwidth]{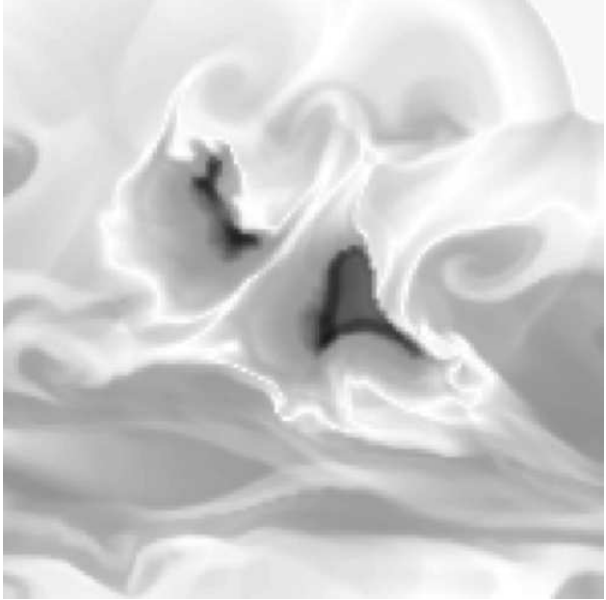}
    \caption{Shocks induced by the passage of the jet are seen sweeping 
     through clouds embedded in the jet cocoon. The grayscale is reveresed 
      at low levels so as to show the jet. Material which has been ablated 
      from the clouds adds to the pressure of the cocoon.}
    \label{fig:1}
\end{figure}

The evolution of the jet itself depends critically on the filling factor of
the clouds. With low filling factor, the jet is bent or diverted by
interaction with clouds, but is not entirely disrupted. In these cases, the
evolution is like the jittering ``dentist's drill'' propagation envisioned
by Sheuer (1982). However, above a certain critical filling factor, the jet
becomes ``frustrated'', and is broken up into a series of channels
reminiscent of a river delta (see Figure 2). In this case, the evolution of
the whole cocoon is more like a cloudy stellar wind bubble. We should
emphasize that jet shown in figure 2 is highly supersonic; its Mach number
is 130. However, the jet kinetic energy is rapidly thermalised so that the 
pressure in the cocoon is very high; $\sim M^2$ times the background pressure.

\begin{figure}
   \leavevmode
    \includegraphics[width=\columnwidth]{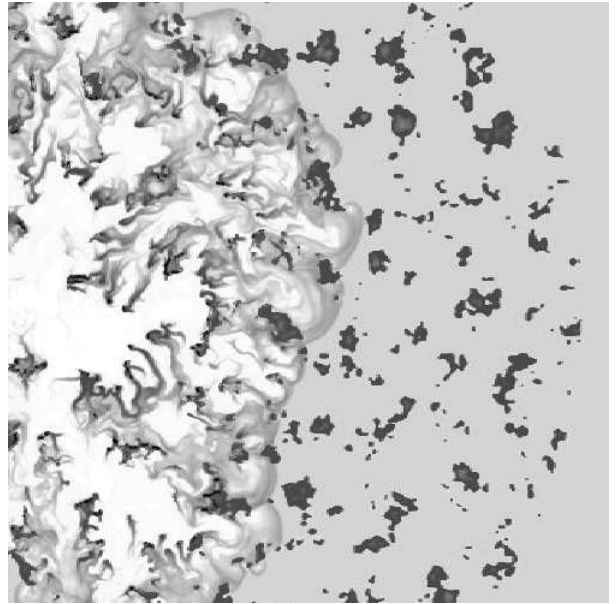}
    \caption{The later evolution of a 2-D  jet surrounded by a set of dense 
             clouds. Above a critical filling factor, the jet is disrupted and 
             ``percolates'' through the channels which it has been able to 
             open up.}
    \label{fig:2}
\end{figure}

\subsection{Hi-Z Radio Sources}
The theory and models discussed above suggest that, in their early phases, 
the jet and circum-nuclear ISM interact very strongly, and that the early 
evolution of the NLR is characterized by the presence of strong radiative 
shocks. Under such circumstances the ``auto-ionizing'' fast shock 
model of \scite{Dopita95} and \scite{Dopita96} would provide the dominant 
source of excitation. 

If this is true in the case of the GPS/CSS sources, it will apply even more
strongly to the gas-rich strong radio sources which are found in abundance in
the high-redshift universe. Such galaxies present us with a statistically 
significant sample in which to study UV line ratio behaviour and so investigate 
the fraction of the NLR emission produced by shocks, and the fraction by
photoionization. The study by Best {\em et al.} (2000a,b) revealed an 
extraordinary result for powerful 3C radio galaxies with $z\sim 1.$ They find 
that both the UV line profiles and the UV\ line ratio diagnostics imply that, 
when the scale of the radio lobes is such that they are still able to interact 
with the gas in the vicinity of the galaxy, they are predominantly 
shock-excited,  but when the lobe has burst out into intergalactic space, the 
ionised gas left behind is predominantly photoionized. The ratio of fluxes in 
the different classes of source suggests that the energy flux in the UV 
radiation field is about 1/3 of the energy  flux in the jets. Thus, both shocks 
and photoionization are important in the overall evolution of radio galaxies. 
This result, confirmed by Inskip {\em et al.} (2002), proves that that 
the properties of the radio jet are intimately connected with the central engine.

Very distant radio galaxies have been recently studied by De Breuck (2000).
He finds that diagnostic diagrams involving C~IV, He~II and C~III] fit to the
pure photoionization models, but that the observed C~II]/C~III] requires there
to be a high-velocity shock present. He argues that composite models would
be required to give a self-consistent description of all the line ratios, and
that these may require a mix of different physical conditions as well.

Such sources are uniquely associated with massive gas-rich multi-$L_*$ galaxies 
in the early universe ($<2-3$Gyr). They display a strong ``alignment effect'',
with regions of very high star formation rate ($>1000~\rm M_\odot~yr^{-1}$), 
and emission line gas having the spectral characteristics of the NLR extended
along the direction of the steep-spectrum radio lobes. In this objects,
the radio jet appears to be driving strong shocks into the galaxian ISM 
(evidenced by extensive Ly-$\alpha$ haloes; \pcite{Reuland03}) which in turn
triggers enormous star formation in the surrounding cocoon.
A fine example of such a source is provided by the $z\sim 3.8$ radio galaxy 
4C 41.17  which has recently been studied in detail by \scite{Bicknell00}. 
This object consists of a powerful ``double-double'' radio source embedded in a 
$190 \times 130$ kpc Ly$\alpha$ halo \scite{Reuland03} and shows
strong evidence for jet-induced star formation at $3000~\rm M_\odot~yr^{-1}$
associated with the inner radio jet.  This is apparently induced by the strong
dynamical interaction of the inner jet with the shocked and compressed gas in the 
wall of the cocoon created by the passage of the outer jet. Shock-induced
starformation in jet walls was proposed in the context of Seyfert galaxies by
\scite{Steffen97}. In 4C 41.17, the outer jet also appears to have 
induced a large-scale outflow with velocities in excess of 500 km~s$^{-1}$ 
in the line-emitting gaseous halo. Thus we may be seeing the ``end of the 
beginning'' in which the central supermassive black hole has finally become
large enough to drive the whole accreting envolope of gas into outflow, 
triggering a last and spectacular burst of star formation in the process.

\subsection{Seyferts}

Seyferts are radio-quiet, and most appear to be dominated by photoionization 
by the central object.The dynamical signatures of strong shocks are apparent 
in only $5-10\%$ of Seyferts (eg. \pcite{Whittle96}), and these tend to be more 
radio-luminous than the average.  However, clear dynamical evidence exists for 
a relatively strong thermal wind arising from the central part of the accretion
disk (\pcite{Rodriguez02}, \pcite{Zamanov02}

In the gas-rich and cloudy circumnuclear environments, light 
and low-power radio jets are readily disrupted and suffer entrainment from
the surrounding material, and molecular clouds are crushed in the high-pressure 
environment. This is clearly demonstrated by the 2-d hydrodynamic simulations
described  above. However, shock velocities in Seyferts will generally be much 
lower than in GPS sources, and although shocks may be still important in 
shaping the circum-nuclear medium, photoionisation by the central nucleus appears
to be much more important for its excitation.

Because clouds lying in the path of the jet and its surrounding high-pressure 
cocoon are crushed at relatively low velocity, any dust mixed with the cloud 
gas is likely to survive. Assuming that the central source 
produces UV photons with high local ionisation parameter, the dusty ionised
gas is compressed, raising the pressure close to the ionisation front
to match the radiation  pressure in the EUV radiation field \scite{Dopita02}). 
The regulates the apparent ionisation parameter, and ensures that the 
density of the photoionized clouds varies as $R^{-2}$.  Each photoablating cloud 
is surrounded by a coronal medium in which the local ionization parameter 
reflects the ``true'' ionisation parameter delivered by the central 
source, and each has a dusty photo-accelerated radial tail, which may 
be accelerated to vey high velocities ($>1000$km~s$^{-1}$)
(c.f. \pcite{Cecil02}). 

\section{Conclusions}
In this paper we have presented observational and theoretical evidence to 
support the following model for the gross features of the temporal
evolution for the NLR:\\
1. Early on, the radio jet (if  there is one), or else a strong outflow of
thermal gas from the inner accretion disk pushes a strong shock into the
surrounding galactic ISM.\\
2. Dense molecular clouds in the shocked region are crushed by radiative
shocks, but along the jet axis, such clouds may be rapidly ablated and
shredded by non-radiative shocks. In this phase, the optical/UV
spectral signature is one of fast radiative shocks.\\
3. Eventually, the jet escapes from the dense region of the galaxy, and
the pressure driving the shocked cocoon drains away. The wall shocks 
then become first radiative, and finally, momentum conserving.
Shock-induced star formation may eventually occur within them, provided
they become unstable to their self-gravity. The opening angle of the
shocked cocoon provides the ``ionization cone''. This opening angle
tends to increase with time, allowing an evolution from a Seyfert II-like
appearance early on, to predominantly Seyfert I-like or BLR-like spectra
at late times.\\
4. In this phase, the crushed clouds left behind in the cocoon are
photoablated by the strong UV field from the central engine, and the 
ablated ionized gas may be radiatively accelerated in the radial
direction. in this phase the NLR presents the signatures of a 
photoionized plasma, and in Seyfert galaxies the coronal lines may be
strong. \\

\begin{acknowledgments}

M. Dopita acknowledges the support of the ANU and  the Australian 
Research Council through his ARC Australian Federation Fellowship, 
and both he and Ralph Sutherland acknowledge support through their 
ARC Discovery project DP0208445. Research by Bicknell, Saxton 
and Sutherland was supported by an ARC Large Grant A69905341.

\end{acknowledgments}



\end{document}